\documentclass[doublespacing]{elsart}

\usepackage{graphicx}
\usepackage{amssymb}
\usepackage{epsf}
\usepackage{epsfig}
\usepackage{psfig}

\begin{document}
\begin{frontmatter}


\journal{SCES'2001: Version 1}


\title{Josephson Plasma Resonance in Tl$_2$Ba$_2$CaCu$_2$O$_8$ in a Magnetic Field Measured using THz Spectroscopy}

%
%
%
%
%

\author{Verner K. Thorsm{\o}lle},
\author{Richard D. Averitt},
\author{Martin P. Maley},
\author{Lev N. Bulaevskii},
\author{Christian Helm},
\author{Antoinette J. Taylor}

%
 
\address{Los Alamos National Laboratory,
Los Alamos, NM 87545, USA}

%
%
%
%


%
%
%
%



\begin{abstract}

We report the first measurements of the $c$-axis Josephson Plasma Resonance (JPR) in Tl$_2$Ba$_2$CaCu$_2$O$_8$ as a function of temperature with and without a $c$-axis magnetic field using terahertz time-domain spectroscopy in transmission. The JPR is sensitive to the alignment of pancake vortices along the $c$-axis, and is observed to decrease when applying a magnetic field as expected. 

\end{abstract}

%
%



\end{frontmatter}

%
%
%
%
%
%
%

The $c$-axis Josephson Plasma Resonance (JPR) in highly anisotropic layered cuprate superconductors originates from the interlayer tunneling of Cooper pairs. The JPR, $\omega_{pc}$ = c/$\lambda_{c}\sqrt{\epsilon^{c}_{\infty}}$ is directly related to the London penetration depth along the $c$-axis, $\lambda_{c}$, and is thus a fundamental probe of the superconducting state and an excellent tool to study these highly anisotropic systems \cite{ThorsmollePhD}. $\epsilon^{c}_{\infty}$ is the high frequency dielectric constant along the $c$-axis. For $T\ll T_{c}$, the temperature dependence of $\lambda_{c}$ is related to the symmetry of the order parameter. For $T$ close to $T_{c}$, the appearance of the JPR probes the onset of interlayer phase coherence. Furthermore, the JPR spectral width is a measure of the quasiparticle scattering rate. In a magnetic field, the JPR probes the correlation of pancake vortices along the $c$-axis and is a tool to study the $B$-$T$ phase diagram \cite{MartinLev95PRL,Kosugi97PRL}.

The JPR of high-$T_{c}$ superconductors with extreme anisotropy such as bismuth, thallium and mercury based high-$T_{c}$ superconductors lie in the far-infrared and is thus amenable to study using terahertz time-domain spectroscopy (THz-TDS). We have measured the JPR in Tl$_{2}$Ba$_{2}$CaCu$_{2}$O$_{8}$ (Tl-2212) superconducting thin films employing THz-TDS in transmission. The sample configuration in the THz beam and the experimental setup is described in ref. \cite{ThorsmollePhD,ThorsmolleOL}.

Figure \ref{fig: 1} shows the electric field amplitude of the terahertz pulse transmitted through the Tl-2212 film plus substrate in the time-domain at different temperatures in zero magnetic field. At 95 K the onset of $c$-axis coherent tunneling is just perceptible as a slight oscillation following the main pulse. At 50 K the terahertz pulse displays an even stronger oscillation amplitude due to a reduction in scattering. This results in a narrowing of the JPR peak in the frequency-domain (See Figure \ref{fig: 2}). The terahertz pulse at 10 K displays a pronounced ringing lasting for at least 20 ps. These traces reveal how the oscillation frequency increases as the temperature is lowered resulting in a shifting of the JPR to higher frequencies. Figure \ref{fig: 2} shows the JPR in the frequency domain. With increasing temperature, the JPR shifts from 710 GHz at 10 K to $\sim$170 GHz at 99 K corresponding to an increase of the $c$-axis penetration depth from 20.0$\pm$0.6 $\mu$m to 84$\pm$9 $\mu$m. The linewidth of the JPR peak increases with temperature, indicative of an increase in the quasiparticle scattering rate. We have probed the onset of the $c$-axis phase coherence to $\sim$0.95$T_{c}$. The JPR vanishes above $T_{c}$ as expected.

Upon application of a magnetic field along the $c$-axis of 2.5 kG, under the field cooling condition, the JPR drops drastically in frequency, due to the induced disorder of pancake vortices. $C$-axis uncorrelated pancake vortices induced by the magnetic field suppresses the interlayer phase coherence and the JPR \cite{Koshelev}. This is shown in Figure \ref{fig: 3}. Measuring the drop in frequency as a function of temperature allows us to determine the crossover from the vortex solid phase to the vortex liquid phase in the $B$-$T$ phase diagram. With a $c$-axis field of 2.5 kG this crossover occurs at $\sim$70 K. The drop in the JPR frequency is in agreement with theoretical predictions \cite{Koshelev}. The details of the calculations will follow in a subsequent publication.

In summary, we have unambiguously observed the JPR in Tl-2212 using THz-TDS in transmission. The JPR dependence on temperature and applied $c$-axis magnetic field is in agreement with theoretical results \cite{Koshelev}, and is demonstrated to be an excellent tool to study the vortex dynamics in highly anisotropic layered superconductors. 

We would like to thank Superconductivity Technologies Inc., Santa Barbara, California, for providing the Tl-2212 films. This research was supported by the University of California Campus-Laboratory Collaborations and by the Los Alamos Directed Research and Development Program by the US Department of Energy.


Corresponding Author: Los Alamos National Laboratory, Superconductivity Technology Center, MS K 763, NM 87545, USA.	Phone: (505) 665-7365
Fax: (505) 665-8601, Email: vthorsmolle@lanl.gov

\begin{figure}
\caption{\label{fig: 1} Electric field of the terahertz pulse in the time-domain at different temperatures in zero magnetic field.}
\end{figure}

\begin{figure}
\caption{\label{fig: 2} Transmission amplitude versus frequency for Tl-2212 film in zero magnetic field. The JPR broadens and shifts to lower frequencies with increased temperature.}
\end{figure}

\begin{figure}
\caption{\label{fig: 3} JPR frequency versus temperature in Tl-2212 film with and without $c$-axis applied field of 2.5 kG ($B||c$).}
\end{figure}


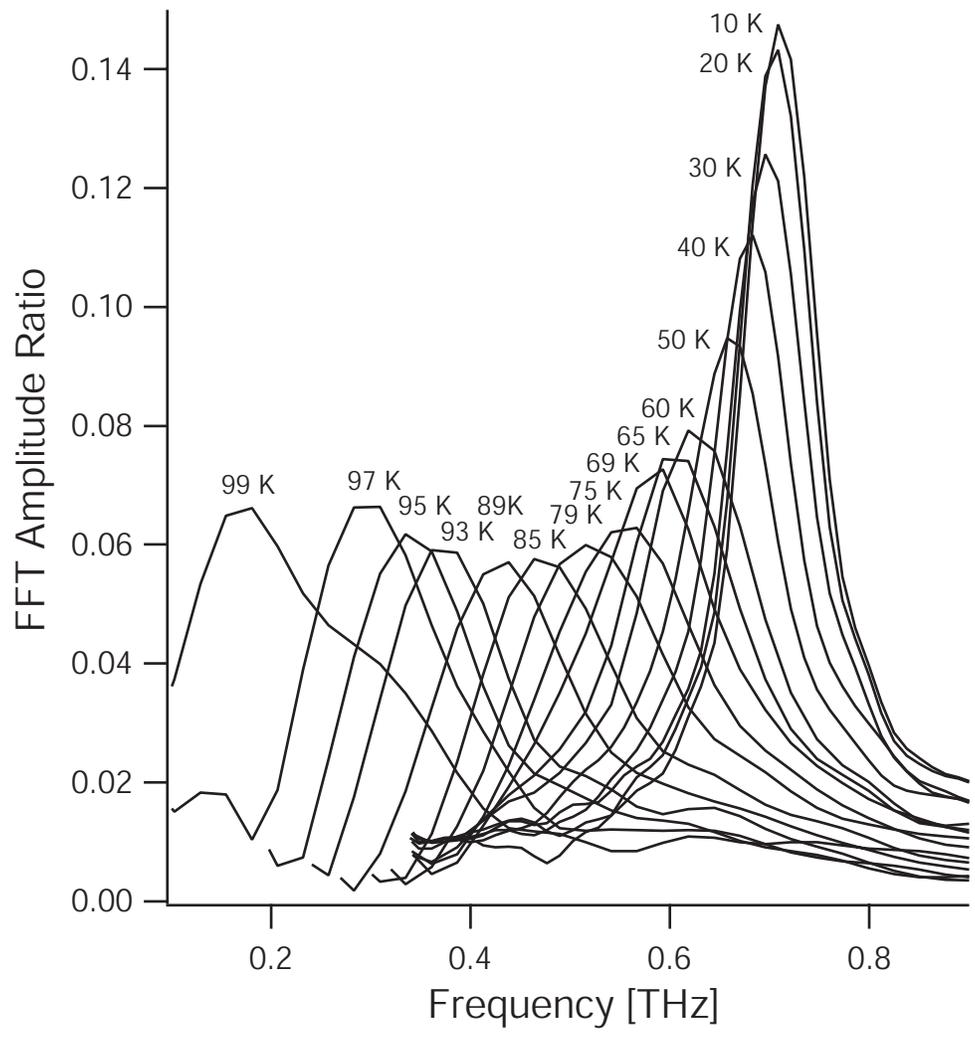
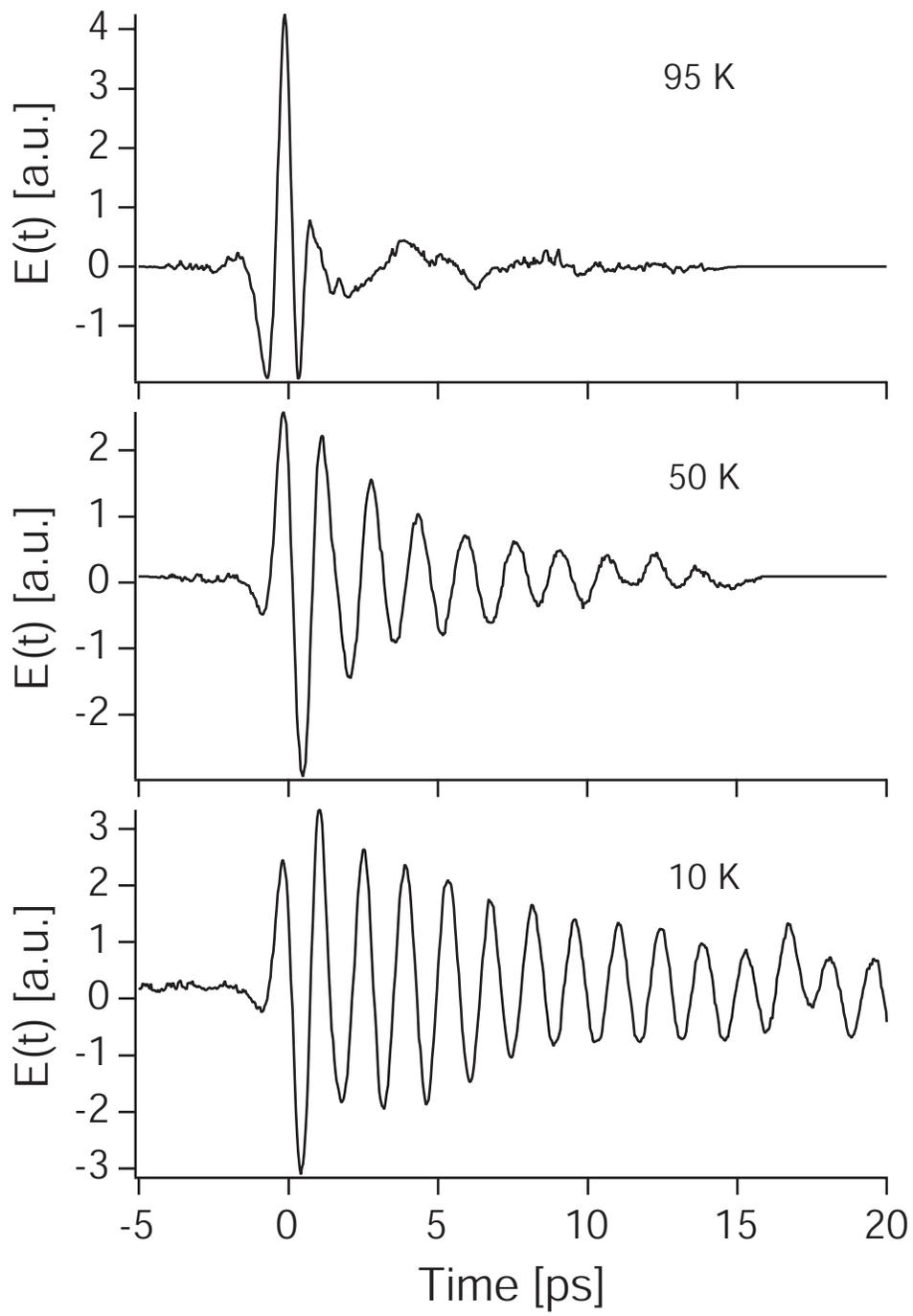
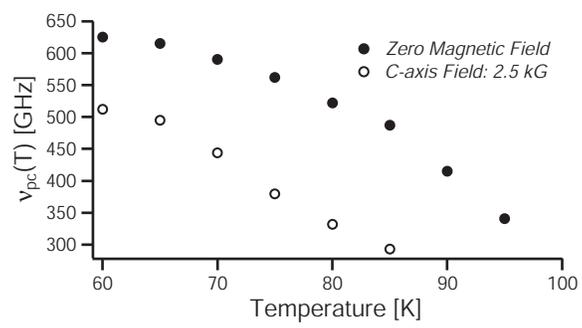
\end{document}